\documentclass{article}

\usepackage{arxiv}

\usepackage[utf8]{inputenc} 
\usepackage[T1]{fontenc}    
\usepackage{lmodern}        
\usepackage{hyperref}       
\usepackage{url}            
\usepackage{booktabs}       
\usepackage{amsfonts}       
\usepackage{nicefrac}       
\usepackage{microtype}      
\usepackage{graphicx}

\title{gratia: An R package for exploring generalized additive models}

\author{
    Gavin L. Simpson
    \thanks{Orcid: 0000-0002-9084-8413; Submitted to the Journal of Open
Source Software}
   \\
    Department of Animal and Veterinary Sciences \\
    Aarhus University \\
  8830 Tjele, Denmark \\
  \texttt{\href{mailto:gavin@anivet.au.dk}{\nolinkurl{gavin@anivet.au.dk}}} \\
  }

\usepackage{color}
\usepackage{fancyvrb}

\DefineVerbatimEnvironment{Highlighting}{Verbatim}{commandchars=\\\{\}}
\usepackage{framed}
\definecolor{shadecolor}{RGB}{248,248,248}
\newenvironment{Shaded}{\begin{snugshade}}{\end{snugshade}}

\newcommand{\AttributeTok}[1]{\textcolor[rgb]{0.13,0.29,0.53}{#1}}

\newcommand{\CommentTok}[1]{\textcolor[rgb]{0.56,0.35,0.01}{\textit{#1}}}

\newcommand{\ConstantTok}[1]{\textcolor[rgb]{0.56,0.35,0.01}{#1}}

\newcommand{\DataTypeTok}[1]{\textcolor[rgb]{0.13,0.29,0.53}{#1}}
\newcommand{\DecValTok}[1]{\textcolor[rgb]{0.00,0.00,0.81}{#1}}

\newcommand{\FloatTok}[1]{\textcolor[rgb]{0.00,0.00,0.81}{#1}}
\newcommand{\FunctionTok}[1]{\textcolor[rgb]{0.13,0.29,0.53}{\textbf{#1}}}

\newcommand{\NormalTok}[1]{#1}

\newcommand{\OtherTok}[1]{\textcolor[rgb]{0.56,0.35,0.01}{#1}}

\newcommand{\SpecialCharTok}[1]{\textcolor[rgb]{0.81,0.36,0.00}{\textbf{#1}}}

\newcommand{\StringTok}[1]{\textcolor[rgb]{0.31,0.60,0.02}{#1}}

\NewDocumentCommand\citeproctext{}{}

\makeatletter
 \let\@cite@ofmt\@firstofone
 \def\@biblabel#1{}
 \def\@cite#1#2{{#1\if@tempswa , #2\fi}}
\makeatother
\newlength{\cslhangindent}
\newlength{\csllabelwidth}
\newenvironment{CSLReferences}[2] 
 {\begin{list}{}{%
  \setlength{\itemindent}{0pt}
  \setlength{\leftmargin}{0pt}
  \setlength{\parsep}{0pt}
  \ifodd #1
   \setlength{\leftmargin}{\cslhangindent}
   \setlength{\itemindent}{-1\cslhangindent}
  \fi
  \setlength{\itemsep}{#2\baselineskip}}}
 {\end{list}}
\usepackage{calc}

\usepackage{amsmath}
\begin{document}
\maketitle

\begin{abstract}

\end{abstract}

\section{Summary}\label{summary}

Generalized additive models (GAMs, Hastie \& Tibshirani, 1990; Wood,
2017) are an extension of the generalized linear model that allows the
effects of covariates to be modelled as smooth functions. GAMs are
increasingly used in many areas of science (e.g. Pedersen, Miller,
Simpson, \& Ross, 2019; Simpson, 2018) because the smooth functions
allow nonlinear relationships between covariates and the response to be
learned from the data through the use of penalized splines. Within the R
(R Core Team, 2024) ecosystem, Simon Wood's \emph{mgcv} package (Wood,
2017) is widely used to fit GAMs and is a \emph{Recommended} package
that ships with R as part of the default install. A growing number of
other R packages build upon \emph{mgcv}, for example as an engine to fit
specialised models not handled by \emph{mgcv} itself (e.g.~\emph{GJMR},
Marra \& Radice, 2023), or to make use of the wide range of splines
available in \emph{mgcv} (e.g.~\emph{brms}, Bürkner, 2017).

The \emph{gratia} package builds upon \emph{mgcv} by providing functions
that make working with GAMs easier. \emph{gratia} takes a \emph{tidy}
approach (Wickham, 2014) providing \emph{ggplot2} (Wickham, 2016)
replacements for \emph{mgcv}'s base graphics-based plots, functions for
model diagnostics and exploration of fitted models, and a family of
functions for drawing samples from the posterior distribution of a
fitted GAM. Additional functionality is provided to facilitate the
teaching and understanding of GAMs.

\section{Generalized additive models}\label{generalized-additive-models}

A GAM has the form \begin{align*}
y_i &\sim    \mathcal{D}(\mu_i, \phi) \\
g(\mu_i) &=  \mathbf{A}_i\boldsymbol{\gamma} + \sum_{j=1} f_j(x_{ji})
\end{align*} where observations \(y_i\) are assumed to be conditionally
distributed \(\mathcal{D}\) with expectation \(\mathbb{E}(y_i) = \mu_i\)
and dispersion \(\phi\). The expectation of \(y_i\) is given by a linear
predictor of strictly parametric terms, whose model matrix is
\(\mathbf{A}_i\) with parameters \(\boldsymbol{\gamma}\), plus a sum of
smooth functions of \(j = 1, \dots, J\) covariates \(f_j()\). \(g()\) is
a link function mapping values on the linear predictor to the scale of
the response.

The smooth functions \(f_j\) are represented in the GAM using penalised
splines, which are themselves formed as weighted sums of basis
functions, \(b_k()\), (De Boor, 2001) e.g. \[
f_j(x_{ij}) = \sum_{k=1}^{K} \beta_{jk} b_{jk}(x_{ij})
\] for a univariate spline. The weights, \(\beta_k\), are model
coefficients to be estimated alongside \(\boldsymbol{\gamma}\). To avoid
overfitting, estimates \(\hat{\beta}_{jk}\) and
\(\hat{\boldsymbol{\gamma}}\) are sought to minimise the penalised
log-likelihood of the model \[
\mathcal{L}(\boldsymbol{\beta}) = \ell(\boldsymbol{\beta}) - \frac{1}{2\phi} \sum_{j} \lambda_{j} \boldsymbol{\beta}^{\mathsf{T}}_j \mathbf{S}_j \boldsymbol{\beta}_j
\] where \(\ell\) is the log likelihood of the data at the parameter
estimates, \(\mathbf{S}_j\) are penalty matrices and \(\lambda_{j}\) are
smoothing parameters associated with each smooth. Note that
\(\boldsymbol{\beta}\) now contains the coefficients
\(\boldsymbol{\gamma}\) and \(\beta_{jk}\).
\(\boldsymbol{\beta}^{\mathsf{T}}_j \mathbf{S}_j \boldsymbol{\beta}_j\)
measures the wiggliness of \(f_j\), which, with the default penalty, is
the integrated squared second derivative of \(f_j\). The smoothing
parameters, \(\boldsymbol{\lambda}\), control the trade-off between fit
to the data and the complexity of the estimated functions.

\begin{figure}

{\centering \includegraphics[width=0.9\linewidth]{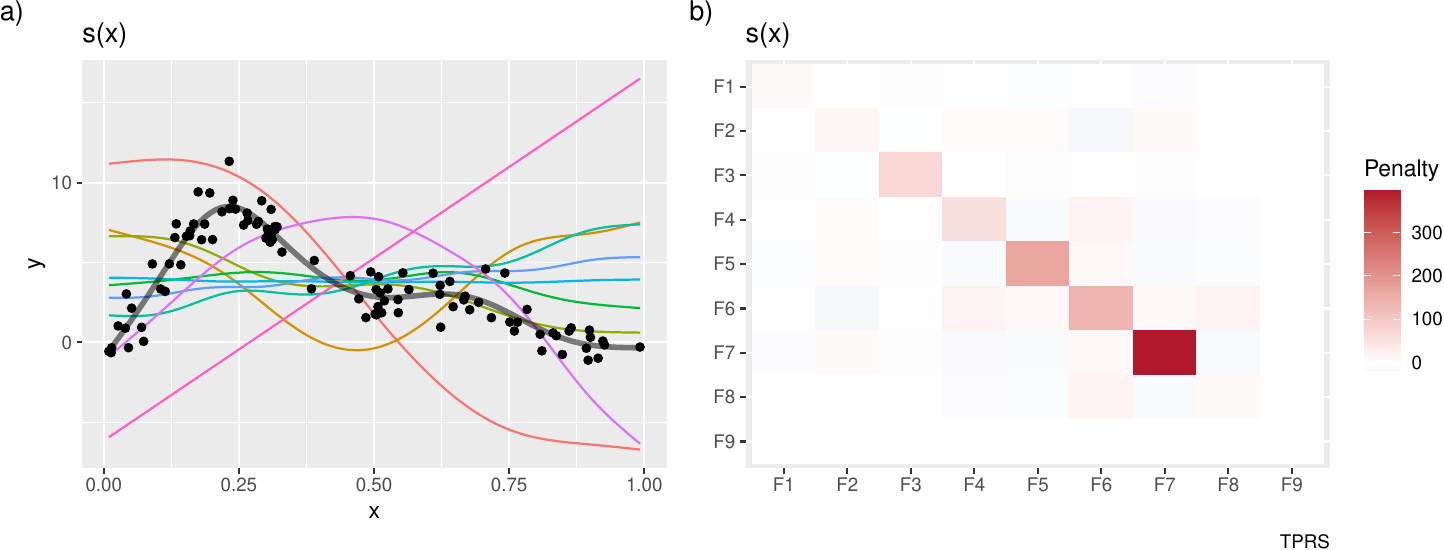} 

}

\caption{\label{fig:basis-funs}Basis functions (a) and associated penalty matrix (b) for a penalised, low rank, thin plate regression spline. a) shows the individual basis functions (thin coloured lines), as well as the data (black points) to which the GAM was fitted. The estimated smooth is shown as the thick grey line. b) shows the penalty matrix for the basis shown in a). Note the 9th basis function (labelled `F9', which is the linear function at the lower left to upper right in a), is not affected by the penalty as it has 0 second derivative everywhere, and hence the resulting penalty for this function is 0.}\label{fig:penalised-spline-basis-and-penalty}
\end{figure}

The default spline created by \emph{mgcv}'s \texttt{s()} is a low rank,
thin plate regression spline (Wood, 2003). Figure \ref{fig:basis-funs},
shows the basis functions for such a spline fitted to data simulated
from the function \[
f = 0.2x^{11}\{10(1 - x)\}^6 + 10(10x)^3(1 - x)^{10} \label{gwf2}
\] with additive Gaussian noise (\(\mu = 0, \sigma = 1\)), and the
associated penalty matrix, prepared using functions from \emph{gratia}.

\section{Statement of need}\label{statement-of-need}

\emph{mgcv} is state-of-the-art software for fitting GAMs and their
extensions to data sets on the order of millions of observations (e.g.
Li \& Wood, 2020; Wood, 2011; Wood, Pya, \& Säfken, 2016). \emph{mgcv}
provides functions for plotting estimated smooth functions, as well as
for producing model diagnostic plots. These functions produce plots
using base graphics, the original plotting system for R. Additionally,
\emph{mgcv} returns fitted GAMs as complex list objects (see
\texttt{?mgcv::gamObject}), the contents of which are not easily used
for downstream analysis without careful study of \emph{mgcv} and its
help pages, plus a good understanding of GAMs themselves. One of the
motivations driving the development of \emph{gratia} was to provide
equivalent plotting capabilities using the \emph{ggplot2} package
(Wickham, 2016). To facilitate this, \emph{gratia} provides functions
for representing the model components as objects using \emph{tidy}
principles, which are suitable for plotting with \emph{ggplot2} or
manipulation by packages in the \emph{tidyverse} (e.g. Wickham,
Cetinkaya-Rundel, \& Grolemund, 2023). This functionality allows for
high-level plotting using the \texttt{draw()} method, as well as easily
customisable plot creation using lower-level functionality.

Taking a Bayesian approach to smoothing with penalized splines
(Kimeldorf \& Wahba, 1970; Silverman, 1985; Wahba, 1983, 1985; see
Miller, 2019 for a summary), it can be shown that GAMs fitted by
\emph{mgcv} are an empirical Bayesian model with an improper
multivariate normal prior on the basis function coeficients. Samples
from the posterior distribution of these models can be used to estimate
the uncertainty in quantities derived from a GAM. This can be invaluable
in applied research, where, for example, a quantity of interest may
arise as an operation on predictions from the model. \emph{gratia}
provides functions for sampling from the posterior distribution of
estimated smooths and from the model as a whole, where sampling can
include the uncertainty in the estimated coefficients
(\texttt{fitted\_samples()}), the sampling uncertainty of the response
(\texttt{predicted\_samples()}), or both
(\texttt{posterior\_samples()}). By default, a Gaussian approximation to
the posterior distribution is used, but a simple Metropolis Hasting
sampler can be substituted (using \texttt{mgcv::gam.mh()}), which has
better performance when the posterior is not well approximated by a
Gaussian approximation.

The teaching of GAMs can benefit from visualisation of the spline basis
functions and associated penalty matrices. \emph{gratia} provides this
functionality via \texttt{basis()} and \texttt{penalty()}, which can be
applied either to a smooth specification
(e.g.~\texttt{s(x,\ z,\ bs\ =\ "ds")}) or to a fitted GAM (see Figure
\ref{fig:basis-funs}). These functions expose functionality already
available in \emph{mgcv}, but supply outputs in a tidy format, which
makes access to these features more intuitive than the original
implementations in \emph{mgcv}. Additional utility functions are
provided, for example: \texttt{model\_constant()}, \texttt{edf()},
\texttt{model\_edf()}, \texttt{overview()}, and \texttt{inv\_link()},
which extract the model intercept term (or terms), the effective degrees
of freedom of individual smooths and the overall model, shows a summary
of the fitted GAM, and extracts the inverse of the link function(s)
used, respectively.

The overall aim of \emph{gratia} is to abstract away some of the
complexity of working with GAMs fitted using \emph{mgcv} to allow
researchers to focus on using and interrogating their model rather than
the technical R programming needed to achieve this. As a result,
\emph{gratia} is increasingly being used by researchers in many fields,
and has, at the time of writing, been cited over 200 times (data from
Google Scholar).

\section{Example usage}\label{example-usage}

In this short example, I illustrate a few of the features of
\emph{gratia} using a data set of sea surface chlorophyll \emph{a}
measurements at a number of locations in the Atlantic Ocean, whose
spatial locations are given as geographical coordinates (\texttt{lat}
and \texttt{lon}), plus two additional covariates; \texttt{bathy}, the
depth of the ocean, in metres, at the sampling location, and
\texttt{jul.day}, the day of the year in which the observation was made.
These data are in the \texttt{chl} dataset provided by the \emph{gamair}
package accompanying Wood (2017).

The packages required for this example are loaded, as is the data set
\texttt{chl} with

\begin{Shaded}
\begin{Highlighting}[]
\NormalTok{pkgs }\OtherTok{\textless{}{-}} \FunctionTok{c}\NormalTok{(}\StringTok{"mgcv"}\NormalTok{, }\StringTok{"gamair"}\NormalTok{, }\StringTok{"gratia"}\NormalTok{, }\StringTok{"ggplot2"}\NormalTok{, }\StringTok{"dplyr"}\NormalTok{, }\StringTok{"ggdist"}\NormalTok{)}
\NormalTok{loaded }\OtherTok{\textless{}{-}} \FunctionTok{vapply}\NormalTok{(pkgs, library, }\FunctionTok{logical}\NormalTok{(}\DecValTok{1}\DataTypeTok{L}\NormalTok{), }\AttributeTok{logical.return =} \ConstantTok{TRUE}\NormalTok{,}
  \AttributeTok{character.only =} \ConstantTok{TRUE}\NormalTok{)}
\FunctionTok{data}\NormalTok{(chl, }\AttributeTok{package =} \StringTok{"gamair"}\NormalTok{)}
\end{Highlighting}
\end{Shaded}

A simple GAM for these data is to model the response (\texttt{chl}) with
a spatial smooth of latitude (\texttt{lat}) and longitude (\texttt{lon})
as covariates. Here, I use a spline on the sphere (SOS) smoother built
using a Duchon spline with second order derivative penalty (Duchon,
1977). Additional terms included in the linear predictor are a smooth of
the day of year of sample collection (\texttt{jul.day}) and a smooth of
ocean depth (\texttt{bath}). The response is assumed to be conditionally
distributed Tweedie, with the power parameter (\(p\)) of the
distribution estimated during fitting. Model coefficients and smoothing
parameters are estimated using restricted maximum likelihood (Wood,
2011)

\begin{Shaded}
\begin{Highlighting}[]
\NormalTok{ctrl }\OtherTok{\textless{}{-}} \FunctionTok{gam.control}\NormalTok{(}\AttributeTok{nthreads =} \DecValTok{10}\NormalTok{)}
\NormalTok{m1 }\OtherTok{\textless{}{-}} \FunctionTok{gam}\NormalTok{(}
\NormalTok{  chl }\SpecialCharTok{\textasciitilde{}} \FunctionTok{s}\NormalTok{(lat, lon, }\AttributeTok{bs =} \StringTok{"sos"}\NormalTok{, }\AttributeTok{m =} \SpecialCharTok{{-}}\DecValTok{1}\NormalTok{, }\AttributeTok{k =} \DecValTok{150}\NormalTok{) }\SpecialCharTok{+}
    \FunctionTok{s}\NormalTok{(jul.day, }\AttributeTok{bs =} \StringTok{"cr"}\NormalTok{, }\AttributeTok{k =} \DecValTok{20}\NormalTok{) }\SpecialCharTok{+}
    \FunctionTok{s}\NormalTok{(bath, }\AttributeTok{k =} \DecValTok{10}\NormalTok{),}
  \AttributeTok{data =}\NormalTok{ chl, }\AttributeTok{method =} \StringTok{"REML"}\NormalTok{, }\AttributeTok{control =}\NormalTok{ ctrl, }\AttributeTok{family =} \FunctionTok{tw}\NormalTok{()}
\NormalTok{)}
\end{Highlighting}
\end{Shaded}

Model diagnostic plots can be produced using \texttt{appraise()}, which
by default produces four plots: i) a QQ plot of model residuals, with
theoretical quantiles and reference bands generated following Augustin,
Sauleau, \& Wood (2012), ii) a plot of residuals (deviance residuals are
the default) against linear predictor values, iii) a histogram of
residuals, and iv) a plot of observed versus fitted values. Model
diagnostic plots for the model, with simulated residuals-based reference
bands on the QQ plot, are produced with

\begin{figure}

{\centering \includegraphics[width=0.9\linewidth]{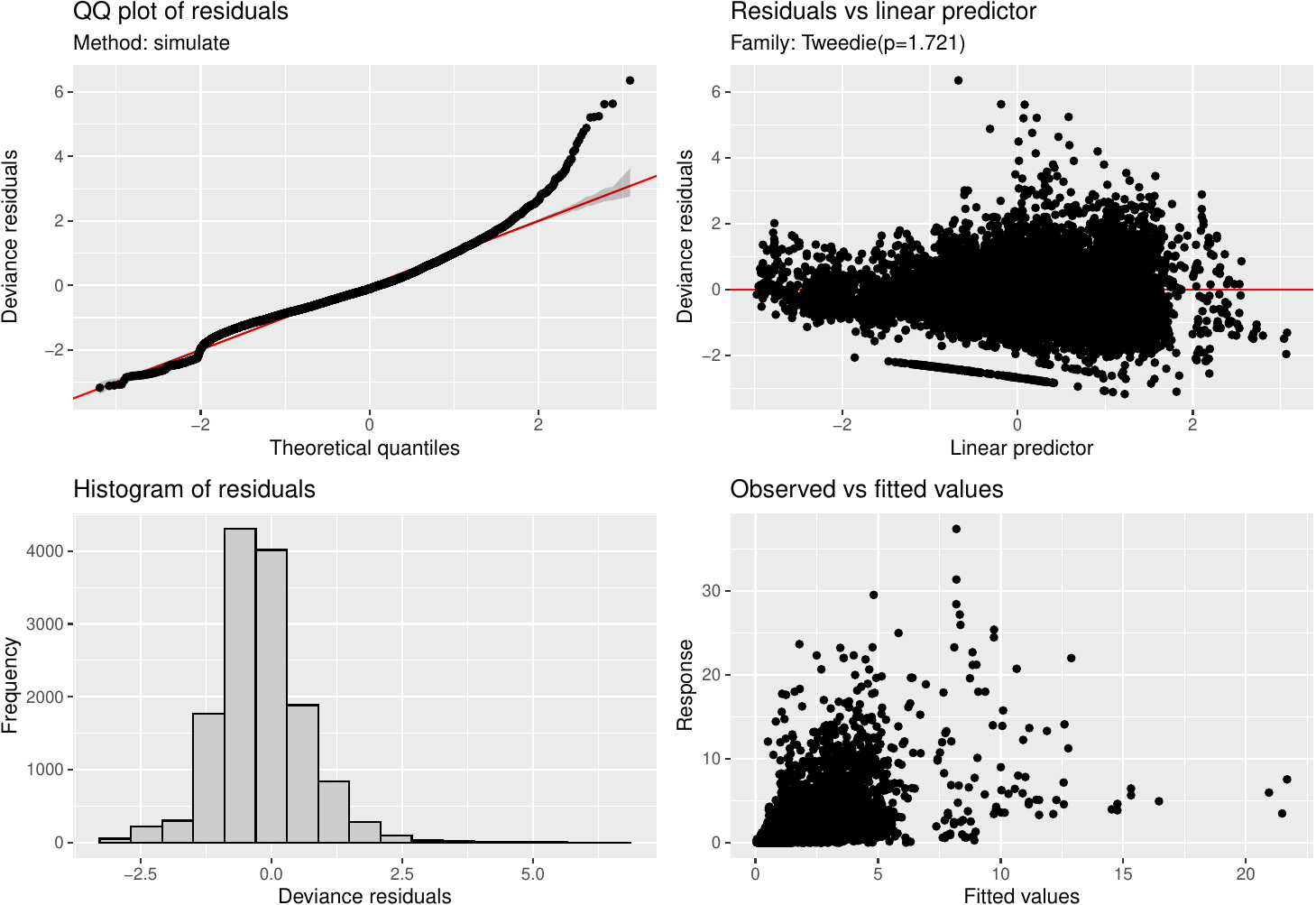} 

}

\caption{Model diagnostic plots for the GAM fitted to the ocean chlorophyll \textit{a} data produced by the \texttt{appraise()} function. The four plots produced are: i) a QQ plot of model residuals, with theoretical quantiles and reference bands generated following Augustin, Sauleau, \& Wood (2012) (upper left), ii) a plot of residuals (deviance residuals are default) against linear predictor values (upper right), iii) a histogram of deviance residuals (lower left), and iv) a plot of observed versus fitted values (lower right)}\label{fig:m1-appraise}
\end{figure}

\begin{Shaded}
\begin{Highlighting}[]
\FunctionTok{appraise}\NormalTok{(m1, }\AttributeTok{method =} \StringTok{"simulate"}\NormalTok{)}
\end{Highlighting}
\end{Shaded}

which show significant heteroscedasticity and departure from the
condtional distribution of the response given the model (Figure
\ref{fig:m1-appraise}).

The problems with the model aparent in the diagnostics plots are
probably due to important controls on chlorophyll \emph{a} missing from
the covariates available in the example data. However, the original
model assumed constant values for the scale, \(\varphi\), and the power
parameter \(p\), which may be too inflexible given the absence of
important effects in the model. A distributional GAM, where linear
predictors for all distributional parameters, may improve the model
diagnostics.

A distributional GAM for \(\mathcal{D}\) Tweedie, with linear predictors
for \(\mu\), \(p\), and \(\varphi\) is fitted below using \emph{mgcv}'s
\texttt{twlss()} family

\begin{Shaded}
\begin{Highlighting}[]
\NormalTok{m2 }\OtherTok{\textless{}{-}} \FunctionTok{gam}\NormalTok{(}
  \FunctionTok{list}\NormalTok{(}
\NormalTok{    chl }\SpecialCharTok{\textasciitilde{}} \FunctionTok{s}\NormalTok{(lat, lon, }\AttributeTok{bs =} \StringTok{"sos"}\NormalTok{, }\AttributeTok{m =} \SpecialCharTok{{-}}\DecValTok{1}\NormalTok{, }\AttributeTok{k =} \DecValTok{150}\NormalTok{) }\SpecialCharTok{+} \CommentTok{\# location}
      \FunctionTok{s}\NormalTok{(jul.day, }\AttributeTok{bs =} \StringTok{"cr"}\NormalTok{, }\AttributeTok{k =} \DecValTok{20}\NormalTok{) }\SpecialCharTok{+}
      \FunctionTok{s}\NormalTok{(bath, }\AttributeTok{k =} \DecValTok{10}\NormalTok{),}
    \SpecialCharTok{\textasciitilde{}} \FunctionTok{s}\NormalTok{(lat, lon, }\AttributeTok{bs =} \StringTok{"sos"}\NormalTok{, }\AttributeTok{m =} \SpecialCharTok{{-}}\DecValTok{1}\NormalTok{, }\AttributeTok{k =} \DecValTok{100}\NormalTok{) }\SpecialCharTok{+}     \CommentTok{\# power}
      \FunctionTok{s}\NormalTok{(jul.day, }\AttributeTok{bs =} \StringTok{"cr"}\NormalTok{, }\AttributeTok{k =} \DecValTok{20}\NormalTok{) }\SpecialCharTok{+}
      \FunctionTok{s}\NormalTok{(bath, }\AttributeTok{k =} \DecValTok{10}\NormalTok{),}
    \SpecialCharTok{\textasciitilde{}} \FunctionTok{s}\NormalTok{(lat, lon, }\AttributeTok{bs =} \StringTok{"sos"}\NormalTok{, }\AttributeTok{m =} \SpecialCharTok{{-}}\DecValTok{1}\NormalTok{, }\AttributeTok{k =} \DecValTok{100}\NormalTok{) }\SpecialCharTok{+}     \CommentTok{\# scale}
      \FunctionTok{s}\NormalTok{(jul.day, }\AttributeTok{bs =} \StringTok{"cr"}\NormalTok{, }\AttributeTok{k =} \DecValTok{20}\NormalTok{) }\SpecialCharTok{+}
      \FunctionTok{s}\NormalTok{(bath, }\AttributeTok{k =} \DecValTok{10}\NormalTok{)),}
  \AttributeTok{data =}\NormalTok{ chl, }\AttributeTok{method =} \StringTok{"REML"}\NormalTok{, }\AttributeTok{control =}\NormalTok{ ctrl, }\AttributeTok{family =} \FunctionTok{twlss}\NormalTok{()}
\NormalTok{)}
\end{Highlighting}
\end{Shaded}

\begin{figure}

{\centering \includegraphics[width=0.9\linewidth]{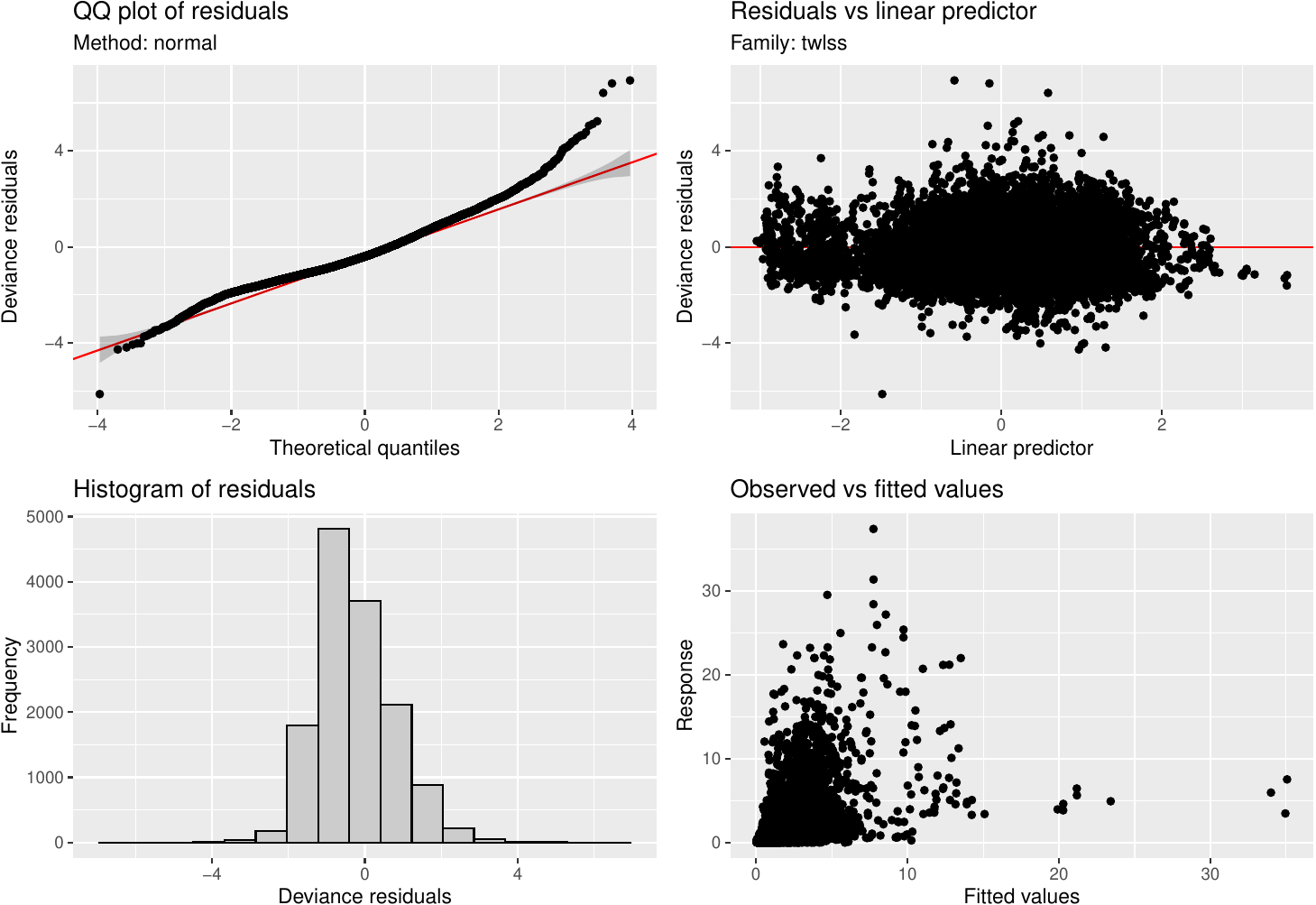} 

}

\caption{Model diagnostic plots for the distributional GAM fitted to the ocean chlorophyll \textit{a} data produced by the \texttt{appraise()} function. Refer to the caption for Figure \ref{fig:m1-appraise} for a description of the plots shown.}\label{fig:m2-appraise}
\end{figure}

This model has much better model diagnostics although some large
residuals remain (Figure \ref{fig:m2-appraise}). Note that the QQ plot
uses theoretical quantiles from a standard normal distribution as the
simulation-based values are not currently available in \emph{mgcv} or
\emph{gratia} for some of the distributional families, including the
\texttt{twlss()} family, and as such, the reference bands may not be
appropriate.

\emph{gratia} can handle distributional GAMs fitted with \emph{mgcv} and
also \emph{GJRM}'s \texttt{gamlss()}. Below, the estimated smooths from
\texttt{m2} are plotted using \texttt{draw()}

\begin{Shaded}
\begin{Highlighting}[]
\NormalTok{crs }\OtherTok{\textless{}{-}} \StringTok{"+proj=ortho +lat\_0=20 +lon\_0={-}40"}
\FunctionTok{draw}\NormalTok{(m2, }\AttributeTok{crs =}\NormalTok{ crs, }\AttributeTok{default\_crs =} \DecValTok{4326}\NormalTok{, }\AttributeTok{dist =} \FloatTok{0.05}\NormalTok{, }\AttributeTok{rug =} \ConstantTok{FALSE}\NormalTok{)}
\end{Highlighting}
\end{Shaded}

\begin{figure}[t!]

{\centering \includegraphics[width=0.9\linewidth]{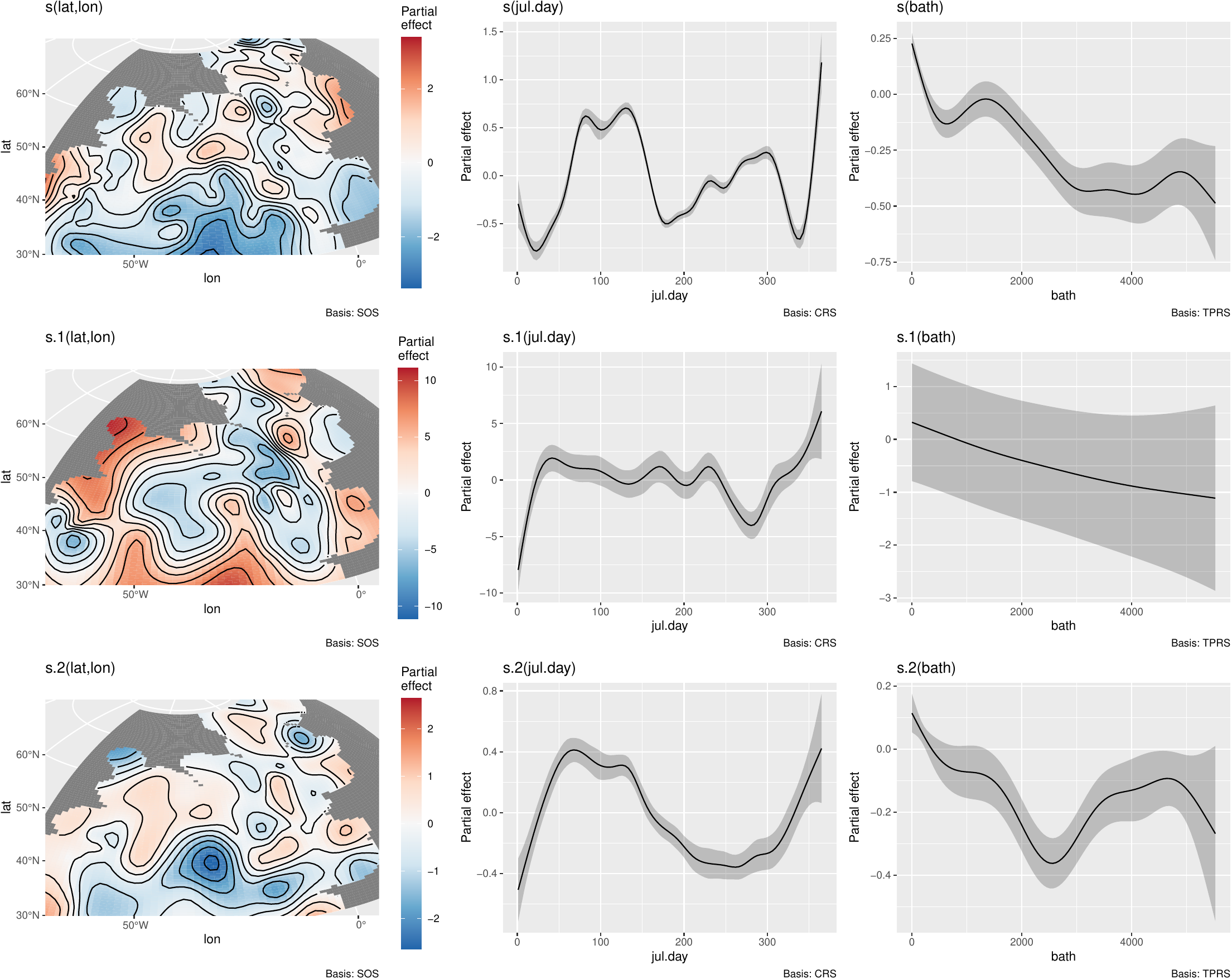} 

}

\caption{Estimated smooth functions for the distributional GAM, \texttt{m2}, fitted to the ocean chlorophyll \textit{a} data. The first row of plots is for the linear predictor of the conditional mean chlorophyll \textit{a}, while the second and third rows are for the conditional power parameter and conditional scale, respectively. The shaded ribbons are 95\% Bayesian credible intervals.}\label{fig:m2-draw}
\end{figure}

Here, we see a specialised plot drawn for spline-on-the-sphere smooths
\(f(\mathtt{lat}_i,\mathtt{lon}_i)\) (Figure \ref{fig:m2-draw}), which
uses \texttt{ggplot2::coord\_sf()} and functionality from the \emph{sf}
package (Pebesma, 2018; Pebesma \& Bivand, 2023) to visualise the smooth
via an orthographic projection.

If the provided plots are insufficient for users' needs, lower-level
functionality is provided by \emph{gratia} to facilitate bespoke
plotting with \emph{ggplot2}. For example, to evaluate the SOS smooth at
a grid (50x50) of values over the range of the covariates, we use
\texttt{smooth\_estimates()} and add a Bayesian credible interval with
\texttt{add\_confint()}:

\begin{Shaded}
\begin{Highlighting}[]
\FunctionTok{smooth\_estimates}\NormalTok{(m2, }\AttributeTok{select =} \StringTok{"s(lat,lon)"}\NormalTok{, }\AttributeTok{n =} \DecValTok{50}\NormalTok{) }\SpecialCharTok{|\textgreater{}}
  \FunctionTok{add\_confint}\NormalTok{()}
\end{Highlighting}
\end{Shaded}

This returns a data frame of the requested values, which is easily
plotted using \texttt{ggplot()}.

\subsection{Posterior sampling}\label{posterior-sampling}

Perhaps we are interested in the average expected chlorophyll \emph{a}
between 40--50 degrees N and 40--50 degrees W. It would be quite a
simple matter to calculate this value from the fitted model: we first
create a slice through the data for the spatial locations were are
interested in using the \texttt{data\_slice()} function, which ensures
that \texttt{ds} contains everything we need to predict from the fitted
model

\begin{Shaded}
\begin{Highlighting}[]
\NormalTok{ds }\OtherTok{\textless{}{-}} \FunctionTok{data\_slice}\NormalTok{(m2,}
  \AttributeTok{lat =} \FunctionTok{evenly}\NormalTok{(lat, }\AttributeTok{lower =} \DecValTok{40}\NormalTok{, }\AttributeTok{upper =} \DecValTok{50}\NormalTok{, }\AttributeTok{by =} \FloatTok{0.5}\NormalTok{),}
  \AttributeTok{lon =} \FunctionTok{evenly}\NormalTok{(lon, }\AttributeTok{lower =} \SpecialCharTok{{-}}\DecValTok{50}\NormalTok{, }\AttributeTok{upper =} \SpecialCharTok{{-}}\DecValTok{40}\NormalTok{, }\AttributeTok{by =} \FloatTok{0.5}\NormalTok{)}
\NormalTok{)}
\end{Highlighting}
\end{Shaded}

Next, \texttt{fitted\_values()} returns the predicted values at the
specified locations. I only include the spatial effects, excluding the
effects of ocean depth and day of year:

\begin{Shaded}
\begin{Highlighting}[]
\NormalTok{use }\OtherTok{\textless{}{-}} \FunctionTok{c}\NormalTok{(}\StringTok{"(Intercept)"}\NormalTok{, }\StringTok{"s(lat,lon)"}\NormalTok{) }\CommentTok{\# , "s.1(lat,lon)", "s.2(lat,lon)")}
\NormalTok{fv }\OtherTok{\textless{}{-}} \FunctionTok{fitted\_values}\NormalTok{(m2, }\AttributeTok{data =}\NormalTok{ ds, }\AttributeTok{terms =}\NormalTok{ use) }\CommentTok{\# predict}
\end{Highlighting}
\end{Shaded}

Finally, I summarise the predictions for the location parameter to yield
the average of the predicted values

\begin{Shaded}
\begin{Highlighting}[]
\NormalTok{fv }\SpecialCharTok{|\textgreater{}}
  \FunctionTok{filter}\NormalTok{(.parameter }\SpecialCharTok{==} \StringTok{"location"}\NormalTok{) }\SpecialCharTok{|\textgreater{}}
  \FunctionTok{summarise}\NormalTok{(}\AttributeTok{chl\_a =} \FunctionTok{mean}\NormalTok{(.fitted))}
\end{Highlighting}
\end{Shaded}

\begin{verbatim}
## # A tibble: 1 x 1
##   chl_a
##   <dbl>
## 1  1.07
\end{verbatim}

While this is an acceptable answer to the question, it lacks an
uncertainty estimate. This is where posterior sampling is useful. With a
small modification of the above code and a little data wrangling, we can
produce an uncertainty estimate ,using \texttt{fitted\_samples()} to
generate posterior draws of the expected chlorophyll \emph{a}:

\begin{Shaded}
\begin{Highlighting}[]
\NormalTok{fs }\OtherTok{\textless{}{-}} \FunctionTok{fitted\_samples}\NormalTok{(m2,     }\CommentTok{\# model}
  \AttributeTok{data =}\NormalTok{ ds,                 }\CommentTok{\# values of covariates to predict at}
  \AttributeTok{terms =}\NormalTok{ use,               }\CommentTok{\# which terms to include in predictions}
  \AttributeTok{n =} \DecValTok{10000}\NormalTok{,                 }\CommentTok{\# number of posterior draws}
  \AttributeTok{method =} \StringTok{"gaussian"}\NormalTok{,       }\CommentTok{\# Gaussian approximation to the posterior}
  \AttributeTok{unconditional =} \ConstantTok{TRUE}\NormalTok{,      }\CommentTok{\# incl uncertainty for estimating lambda}
  \AttributeTok{n\_cores =} \DecValTok{4}\NormalTok{,               }\CommentTok{\# how many CPU cores to compute MVN samples}
  \AttributeTok{seed =} \DecValTok{342}\NormalTok{)                }\CommentTok{\# set the random number seed, used internally}
\end{Highlighting}
\end{Shaded}

The posterior draws can then be summarised as before, except now the
average chlorophyll \emph{a} is calculated separately for each posterior
draw (\texttt{.draw})

\begin{Shaded}
\begin{Highlighting}[]
\NormalTok{fs }\SpecialCharTok{|\textgreater{}}                                 \CommentTok{\# take the posterior draws}
  \FunctionTok{group\_by}\NormalTok{(.draw) }\SpecialCharTok{|\textgreater{}}                  \CommentTok{\# group them by \textasciigrave{}.draw\textasciigrave{}}
  \FunctionTok{summarise}\NormalTok{(}\AttributeTok{chl\_a =} \FunctionTok{mean}\NormalTok{(.fitted)) }\SpecialCharTok{|\textgreater{}} \CommentTok{\# compute mean of fitted chl a}
\NormalTok{  ggdist}\SpecialCharTok{::}\FunctionTok{median\_qi}\NormalTok{()                 }\CommentTok{\# summarise posterior}
\end{Highlighting}
\end{Shaded}

\begin{verbatim}
## # A tibble: 1 x 6
##   chl_a .lower .upper .width .point .interval
##   <dbl>  <dbl>  <dbl>  <dbl> <chr>  <chr>    
## 1  1.07  0.866   1.34   0.95 median qi
\end{verbatim}

The posterior distribution of average chlorophyll \emph{a} is summarised
using \texttt{median\_qi()} from the \emph{ggdist} package (Kay, 2024a,
2024b). While it would be a simple matter to compute the interval with
base R commands, the use of \texttt{median\_qi()} illustrates how
\emph{gratia} tries to interact with other packages.

\section*{References}\label{references}
\addcontentsline{toc}{section}{References}

\phantomsection\label{refs}

\bibliographystyle{unsrt}
\bibliography{paper.bib}

\end{document}